\documentclass[aps,prl,twocolumn,groupedaddress,showpacs]{revtex4}
\usepackage{amssymb}
\usepackage{amsmath, amsthm}

\begin{document}


\title{Comments on ``Solar System constraints to general $f(R)$ 
gravity''}


\author{Valerio Faraoni}
\email[]{vfaraoni@ubishops.ca}
\author{Nicolas Lanahan-Tremblay}
\email[]{ntremblay05@ubishops.ca}
\affiliation{Physics Department, Bishop's University, 2600 
College St., Sherbrooke, Qu\'{e}bec, Canada J1M~0C8}


\date{\today}

\begin{abstract}
We comment on, and complete, the analysis of the weak-field 
limit of metric $f(R)$ gravity in Ref.~\cite{CSE}. 
\end{abstract}
\pacs{04.25.Nx, 04.50.+h}

\maketitle



Recently, Chiba, Smith, and Erickcek \cite{CSE} discussed the 
weak-field limit of $f(R)$ gravity in the metric formalism. 
These modified gravity theories are widely used as alternatives 
to dark energy models to explain the present acceleration of the 
universe.  The authors of Ref.~\cite{CSE} consider the 
weak-field limit of 
metric $f(R) $ gravity without resorting to the equivalence with 
scalar-tensor theory. This direct approach to the weak-field 
limit is highly desirable 
and convincing, as complete physical 
equivalence between 
scalar-tensor and $f(R)$ gravity was questioned {\em a priori}. 
Chiba, Smith, and Erickcek perform an expansion around a de 
Sitter background of constant Ricci curvature $R_0$, with a 
slight metric deviation due to a spherically symmetric 
perturbation describing a star-like  object.

Chiba, Smith, and Erickcek's analysis relies on three 
assumptions:
\begin{enumerate}
\item  $f(R)$ is analytical  at $ 
R_0$.

\item ~~$mr<<1$, where $m$ is the effective mass of 
the scalar field degree of freedom of the theory. This scalar 
field (the Ricci  curvature or, better, the field $f'(R)$, which 
is a dynamical  quantity in the metric 
formalism) must be light and long-ranged. If its range is much 
shorter 
than $\sim 0.2$~mm \cite{EotWash}, this scalar 
is effectively hidden 
from Solar System and terrestrial experiments. In this case, 
this field could not have cosmological effects at late times, 
but could only be used in the very early universe at high 
curvatures, {\em e.g.}, in Starobinsky-like inflation 
\cite{Starobinsky}.

\item For the energy-momentum of a local 
star-like object $P\simeq 0$. The trace of 
a fluid  energy-momentum tensor (in units $c=1$) is 
therefore $T_1=-\rho+3P\simeq -\rho$.

\end{enumerate}

Not all $f(R)$ functions satisfy assumption~1), as noted in 
\cite{CSE}. The validity of assumption~2) about the scalar field  
$R$ being light has already been discussed in several papers 
\cite{HuSawickietc}. Due to the chameleon effect 
well-known in quintessence models \cite{chameleon}, the 
effective scalar has  a mass and range that depend on the 
background density and curvature. As  a result, this field can 
be short-ranged at cosmological curvatures,  
evading the weak-field limit constraints studied by \cite{CSE} 
and still causing the cosmic acceleration. $f(R)$  models 
with the chameleon effect have 
been discussed extensively \cite{HuSawickietc}, and we will not 
comment 
further on them here. Instead, we would like to refine the 
analysis of Chiba, Smith, and  Erickcek of the models that {\em 
do} satisfy assumptions 1)--3). For these models, we fully agree 
with their conclusions but we partially disagree on the 
following.

Eq.~(10) of \cite{CSE} for the Ricci scalar perturbation $R_1$ 
 is 
\begin{equation} \label{eqforR1}
\nabla^2R_1 -m^2 R_1=-\frac{\kappa\, \rho}{3f_0''} \;,
\end{equation}
where $m^2$ is an effective mass squared given by (eq.~(11) of 
\cite{CSE})
\begin{equation}  
m^2 = \frac{1}{3} \left( \frac{f_0'}{f_0''} - R_0 
-\frac{3\Box f_0''}{f_0''}\right) \;.
\end{equation}
The last term, proportional to $\Box f_0''$ in $ m^2$ should be 
dropped. Many discussions of this effective mass of $R_1$ in a 
de Sitter background appeared in the literature, 
including propagator calculations and (gauge-independent)  
studies of inhomogeneous and  homogeneous perturbations of de 
Sitter space, and they all find 
\begin{equation}  
m^2 = \frac{1}{3} \left( \frac{f_0'}{f_0''} - R_0 \right) =
\frac{ ( f_0')^2 -2 f_0 f_0'' }{3 f_0' f_0'' } 
\end{equation}
as the effective mass squared
\cite{Olmo,NavarroVanAcoleyen,NunezSolganik, 
FaraoniNadeau,FaraoniPRD2007, mattmodgrav, HuSawicki2papers}, 
where the last 
equality follows from the condition for the existence of de 
Sitter space $6H_0^2 f_0'-f_0=0  $. Chiba, Smith, 
and Erickcek then provide Green functions for the cases $m^2>0$ 
and $m^2<0$, treating both cases as viable in principle. 
However, the case $m^2<0$ corresponds to an unstable de Sitter 
space, which is  ruled out,  in {\em all} models of 
metric $f(R)$ gravity 
\cite{FaraoniNadeau,FaraoniPRD2007, mattmodgrav}. It 
is true, however, that the limit $m \rightarrow 0$ is taken in 
the subsequent discussion of 
\cite{CSE} and this point does not affect their final results.

Chiba, Smith, and Erickcek proceed to solve the weak-field 
equations to obtain the post-Newtonian potentials $\Psi(r)$ and 
$\Phi(r)$ due to a spherically symmetric perturbation of de 
Sitter space appearing in the line element of modified gravity
\begin{eqnarray}
&& ds^2=-\left[ 1 -2 \Psi(r) -H_0^2r^2\right]dt^2+\left[ 1+2 
\Phi(r)  \right.  \nonumber \\
&& \nonumber \\
&&\left. + H_0^2r^2 \right] dr^2
 +r^2\left( d\theta^2 +\sin^2 \, \varphi \right)  \;.
\end{eqnarray}
Neglecting terms of order $H_0r$, the weak-field equations 
become
\begin{equation} \label{24ofCSE}
f_0' \nabla^2 \Psi(r)=\frac{2\kappa\, \rho}{3} -\frac{f_0'}{2}\, 
R_1 
\end{equation}
(eq.~(24) of \cite{CSE}) and
\begin{equation} \label{21ofCSE}
f_0'\left( -\frac{d^2\Psi}{dr^2}+\frac{2}{r}\frac{d\Phi}{dr} 
\right) -\frac{f_0'R_1}{2}+\frac{2f_0''}{r} \frac{dR_1}{dr} 
\simeq 0
\end{equation}
(eq.~(21) of \cite{CSE}). Our last remark concerns the 
solution of 
these equations. Following \cite{CSE}, the solution of 
eq.~(\ref{24ofCSE}) is expressed as 
$\Psi(r)=\Psi_0(r)+\Psi_1(r)$, 
with
\begin{eqnarray}
f_0' \nabla^2 \Psi_0 &= & \frac{2}{3}\, \kappa\, \rho \;, \\
&& \nonumber \\
f_0' \nabla^2 \Psi_1 &= & - \frac{f_0'}{2}\, R_1 \;.   
\label{starstar} 
\end{eqnarray}
The authors of \cite{CSE} obtain 
\begin{equation}
\Psi_0(r)=-\, \frac{\kappa M}{6\pi f_0'} \, \frac{1}{r} 
\end{equation}
outside the star-like object, and eq.~(\ref{starstar}) is 
integrated twice, yielding
\begin{equation}
\Psi_1(r)=\frac{-\kappa\, M r }{48\pi f_0''}-\frac{C_1}{r}+C_2 
\;,
\end{equation}
where $C_{1,2} $ are integration constants. $C_2$ can be set to 
zero as customary in the Newtonian limit. Chiba, Smith and 
Erickcek drop the constant $C_1$ without discussion. If $C_1\neq 
0$, then the potential $\psi(r)$ contains a term that is 
singular at the origin: therefore, a reasonable boundary 
condition  at $r=0$ consists of imposing $C_1=0$. (In any case, 
including a term with $C_1\neq 0$ 
would not change the final result of \cite{CSE} that the PPN 
parameter $\gamma$ is close to $1/2$ instead of unity.)

\begin{acknowledgments}
This work was supported by a Bishop's 
University Research Grant and by the Natural Sciences and 
Engineering Research Council of Canada (NSERC).
\end{acknowledgments}


\begin{thebibliography}{99}

\bibitem{CSE} T. Chiba, T.L. Smith, and A.L. Erickcek, {\em 
Phys. Rev. D} {\bf 75}, 124014 (2007).

\bibitem{EotWash} C.D. Hoyle {\em et al.}, {\em Phys. Rev. 
Lett.} {\bf 86}, 1418 (2001).

\bibitem{Starobinsky} A.A. Starobinsky, {\em Phys. Lett. B} {\bf 
91}, 99 (1980).

\bibitem{HuSawickietc} W. Hu and I. Sawicki, {\em Phys. Rev> D} 
{\bf 76}, 064004 (2007);
Y.-S. Song, H. Peiris, and W. Hu, {\em Phys. Rev. D} {\bf 76}, 
063517 (2007);
A.A. Starobinsky, arXiv:0706.2041;
S.A. Appleby and R.A. Battye, {\em Phys. Lett. B} {\bf 654}, 7 
(2007);
S. Tsujikawa, {\em Phys. Rev. D} {\bf 76}, 023514 (2007); 
arXiv:0709.1391; S. Nojiri and S.D. Odintsov, arXiv:0707.1941; 
{\em Phys. Rev. D} {\bf 74}, 086005 (2006).

\bibitem{chameleon} J. Khoury and A. Weltman, {\em Phys. 
Rev. Lett.} {\bf 93}, 171104 (2004);  {\em 
Phys. Rev. D} {\bf 69}, 044026 (2004).

\bibitem{Olmo} G.J. Olmo, {\em Phys. Rev. D} {\bf 75}, 
023511 (2007).

\bibitem{NavarroVanAcoleyen} 
I. Navarro and K. Van Acoleyen, {\em J. Cosmol. Astropart. 
Phys.} 0702:022 (2007).

\bibitem{NunezSolganik} A. N\'{u}nez and S. Solganik, 
hep-th/0403159.

\bibitem{FaraoniNadeau} V. Faraoni and S. Nadeau, {\em Phys. 
Rev. D} {\bf 72}, 124005 (2005).

\bibitem{FaraoniPRD2007} V. Faraoni, {\em Phys. Rev. D} {\bf 
75}, 067302 (2007);  {\bf  70},  044037 (2004); {\bf  72}, 
061501(R) (2005); {\bf 69}, 123520 (2004).

\bibitem{mattmodgrav} V. Faraoni, {\em Phys. Rev. D} {\bf 74}, 
104017 (2006).

\bibitem{HuSawicki2papers} I. Sawicki and  W. Hu, {\em Phys. 
Rev. D} {\bf 75}, 127502 (2007);
Y.-S. Song, W. Hu, and I Sawicki, {\em Phys. Rev. D} {\bf 75}, 
044004 (2007).

\bibitem{BertottiIessTortora} B. Bertotti, L. Iess, and P. 
Tortora, {\em  Nature} {\bf 425}, 374 (2003).

\end{thebibliography}

\end{document}